\documentclass[a4paper]{PoS}

\def \be  {\begin{equation}}
\def \ee  {\end{equation}}
\def \ee  {\end{equation}}
\def \bea {\begin{eqnarray}}
\def \eea {\end{eqnarray}}

\newcommand{\nn}{\nonumber}

\title{SU($4$) Polyakov linear-sigma model at finite temperature and density}

\ShortTitle{SU($4$) PLSM at finite temperature and density}

\author{Abdel Magied Diab 
\\
        Egyptian Center for Theoretical Physics (ECTP), Modern University for Technology and Information (MTI), 11571 Cairo, Egypt,\\
        World Laboratory for Cosmology And Particle Physics (WLCAPP), 11571 Cairo, Egypt. 
}

\author{Azar I. Ahmadov\\
 Department of Theoretical Physics, Baku State University, Z. Khalilov st. 23, AZ-1148, Baku, Azerbaijan.
}

\author{\speaker{Abdel Nasser Tawfik} \\
        Egyptian Center for Theoretical Physics (ECTP), Modern University for Technology and Information (MTI), 11571 Cairo, Egypt,\\
        World Laboratory for Cosmology And Particle Physics (WLCAPP), 11571 Cairo, Egypt.\\
        E-mail: \email{a.tawfik@eng.mti.edu.eg}
        }

\author{Eiman Abou El Dahab\\
        Faculty of Computer Sciences, Modern University for Technology and Information,
 11671 Cairo, Egypt.
        }
        
\abstract{In mean-field approximation, the SU($4$) Polyakov linear-sigma model (PLSM) is constructed in order to characterize the quark-hadron phase structure in a wide range of temperatures and densities. The chiral condensates  $\sigma_l$, $\sigma_s$ and  $\sigma_c$ for light, strange and charm quarks, respectively, and the deconfinement order-parameters $\phi$ and $\phi^*$ shall  be analysed at finite temperatures and densities. We conclude that, the critical temperatures corresponding to charm condensates are greater than that to strange and light ones, respectively. Thus, the charm condensates are likely not affected by the QCD phase-transition. Furthermore, increasing the chemical potentials decreases the corresponding critical temperatures.
}

\FullConference{38th International Conference on High Energy Physics\\
		3-10 August 2016\\
		Chicago, USA}
		
\begin{document}

\section{Introduction}
\label{Intro}

Quantum chromodynamics (QCD) describes the strong interactions between quarks and gluons. As the temperature  decreases or at low densities, the quarks and gluons become confined and the physical degrees-of-freedom can be dominated by hadrons.  By imposing chiral symmetries, various approaches besides lattice QCD simulations (LQCD) aim at describing the quark-hadron phase transitions in thermal and dense QCD medium, see for instance Refs. \cite{Gasiorowicz,Meissner}.  At vanishing quark masses, the chiral symmetries represent basic properties of the QCD Lagrangian, while at finite current quark masses, the chiral symmetry is explicitly broken \cite{Vafa:1984} and the QCD phase-diagram can be described by the first-principle LQCD \cite{Gross:1973, Politzer:1973} and various QCD-like approaches.  In charactering SU($2$) and SU($3$), the Polyakov Nambu-Jona Lasinio (PNJL) model \cite{Fukushima:2004,Fukushima:2008,Ratti:2005} and the Polyakov linear-sigma model (PLSM) \cite{Lenaghan,TiwariA,TiwariB,Schaefer:2008hk,Walaa:2009a,Kovacs:2006,OURPLSM201560, OURPLSM201561, OURPLSM201562, OURPLSM201563, OURPLSM201564, OURPLSM201565, OURPLSM201566, OURPLSM201567,OURPLSM201568} have been implemented. The SU($4$) Lagrangian has the same structure as that of SU($3$) \cite{Lenaghan}, for instance. All meson fields should be parametrized in terms of $4 \times 4$ instead of $3 \times 3$ matrices and the chiral phase-structures can be studied for mesons having masses up to $\sim3.5$~GeV. The corresponding quark number susceptibilities and correlations can also be analysed.

The present work is organized as follows. We utilize mean-filed approximation to the PLSM with $N_f=4$ quarks flavors at finite temperatures and baryon chemical potentials in section \ref{model}. The thermal dependence of the light, strange and charm quark chiral-condensates, $\sigma_l,\, \sigma_s$, and $\sigma_c$, respectively, and the corresponding deconfinement order-parameters, $\phi$ and $\phi^*$, the Ployakov-loop fields, shall be discussed in section \ref{results}. Section \ref{Conclusion} is devoted to the final conclusions.

\section{SU($4$) Polyakov linear-sigma model \label{model}}

In SU($4$)$_L$ $\times$ SU($4$)$_R$ symmetries, the LSM Lagrangian \cite{Lenaghan} can be constructed as  $\mathcal{L}_{\mathrm{chiral}}=\mathcal{L}_q + \mathcal{L}_m$, where $q\in(u,d,s,c)$, 
\begin{equation}
\mathcal{L}_q = \sum_f \overline{q}_f \left[i\gamma^{\zeta} D_{\zeta}-gT_a(\sigma_a+i \gamma_5 \pi_a)\right] q_f, \label{eq:quarkL}
\end{equation}
and $g$ is the flavor-blind Yukawa coupling \cite{blind}. $\zeta$ is an additional Lorentz index. The pure mesonic part is given as    
\bea
\mathcal{L}_m &=&  \mathrm{Tr}(\partial_{\mu}\Phi^{\dag}\partial^{\mu}\Phi-m^2 \Phi^{\dag} \Phi)-\lambda_1 [\mathrm{Tr}(\Phi^{\dag} \Phi)]^2 -\lambda_2 \mathrm{Tr}(\Phi^{\dag}
\Phi)^2 + \mathrm{Tr}[H(\Phi+\Phi^{\dag})],  \label{lmeson}
\eea
to which an $c$-term, $c[\mathrm{Det}(\Phi)+\mathrm{Det}(\Phi^{\dag})]$, is usually added, where $\Phi$ is a complex $4\times4$ matrix for scalar and pseudoscalar mesons $\sigma_a$ and $\pi_a$, respectively, $\Phi =T_a(\sigma_a+i\pi_a),$ with $T_a=\lambda_a/2$ with $a = 0, \cdots, N_f^2-1$ are the generators of $U(4)$ symmetry group \cite{Lenaghan}. $\lambda_a$ are Gell-Mann matrices, while $\lambda_0 = {\bf \hat{I}}/\sqrt{2}$ \cite{Gell Mann:1960}. 

The Polyakov-loop potentials introduces color degrees-of-freedom and gluon dynamics, where $\phi$ and $\phi^*$, $\mathcal{L}_{\mathrm{PLSM}}=\mathcal{L}_{\mathrm{chiral}} -\mathcal{U} (\phi, \phi^*, T)$ are Polyakov-loop fields. Details about the chiral LSM Lagrangian for $N_f=2$ and $3$ can be found in Refs. \cite{Lenaghan,Schaefer:2008hk,Schaefer:2007d,Sasaki:2013ssdw}. In the present work, we use polynomial logarithmic parametrisation potential for the Polyakov-loop fields, \cite{Sasaki:2013ssdw}
\bea
\frac{\mathbf{\mathcal{U}}_{\mathrm{PolyLog}}(\phi, \phi^*, T)}{T^4} &=&   \frac{-a(T)}{2} \; \phi^* \phi + b(T)\; \ln{\left[1- 6\, \phi^* \phi + 4 \,( \phi^{*3} + \phi^3) - 3 \,( \phi^* \phi)^2 \right]} \nn \\ &+& \frac{c(T)}{2}\, (\phi^{*3} + \phi^3) + d(T)\, ( \phi^* \phi)^2. \label{LogPloy}
\eea 
The various coefficients $a$, $c$, and $d$ have been determined in Ref. \cite{Sasaki:2013ssdw}, $x(T) = [x_0 + x_1 \left(T0/T\right) + x_2 \left(T0/T\right)^2]/[1+x_3 \left(T0/T\right) + x_4 \left(T0/T\right)^2]$ and $b(T)= b_0\, \left(T0/T\right)^{b_1} [1-e^{b_2 \left(T0/T\right)^{b_3}}]$, with $x=(a,\,c,\,d)$. 

In mean-field approximation,  all fields are treated as constants (averages) in space and imaginary time $\tau$. The exchanges between particles and antiparticles shall be expressed as function of temperature ($T$) and chemical potential ($\mu$), for instance. Accordingly, grand-canonical partition function ($\mathcal{Z}$) can be constructed, from which the thermodynamic potential density can be derived $\Omega (T, \mu)= \Omega_{\bar{q}q}(T, \mu) +\mathcal{U}_{\mathrm{PolyLog}} (\phi, \phi^*, T)+ U(\sigma_l, \sigma_s, \sigma_c)=-T\, \ln{\mathcal{Z}}/V$.
The thermodynamic antiquark-quark potential, $\Omega_{\bar{q}q}(T, \mu)$, was introduced in Ref. \cite{Fukushima:2008, Kapusta:2006pm},
\bea
\Omega_{\bar{q}q}(T, \mu)&=& -2T \sum_{f=l, s, c} \int_0^{\infty} \frac{d^3\vec{P}}{(2 \pi)^3} \left\{ \ln \left[ 1+3\left(\phi+\phi^* e^{-\frac{E_f-\mu}{T}}\right)\times e^{-\frac{E_f-\mu}{T}}+e^{-3 \frac{E_f-\mu}{T}}\right] \right. \nonumber \\ 
&& \hspace*{32.2mm} \left.  +\ln \left[ 1+3\left(\phi^*+\phi e^{-\frac{E_f+\mu}{T}}\right)\times e^{-\frac{E_f+\mu}{T}}+e^{-3 \frac{E_f+\mu}{T}}\right] \right\}. \hspace*{8mm} \label{PloykovPLSM}
\eea
The subscripts $l$, $s$, and $c$ refer to degenerate light, strange and charm quarks, respectively. The energy-momentum relation is given as $E_f=(\vec{P}^2+m_f^2)^{1/2}$, with $m_f$ being the flavor mass of light, strange, and charm quark coupled to $g$ \cite{blind}; $m_l = g \sigma_l/2$, $m_s=g \sigma_s/\sqrt{2}$ \cite{Kovacs:2006}, and $m_c=g \sigma_c/\sqrt{2}$. 

The values of the vacuum chiral-condensates are determined from pion, kaon and D-meson decay widths by means of the partially conserved axial-vector current relation (PCAC) \cite{Lenaghan, Walaa:2009a}. At $T=0$, the quark condensates reads $\sigma_{l_0}=f_\pi=92.4~$MeV, $\sigma_{s_0}= (2f_K-f_\pi)/\sqrt{2}= 94.5~$MeV and $\sigma_{c_0} =(2f_D-f_\pi)/\sqrt{2}= 293.87~$MeV. 

By introducing SU($4$)$_L \times$ SU($4$)$_R$ symmetries \cite{Lenaghan}, the purely mesonic potential can be driven in basis of light, strange and charm quark from the mesonic Lagrangian, $\mathcal{L}$. The orthogonal basis transformation from the original ones: $\sigma_0,  \sigma_8$ and  $\sigma_{15}$,  to the light $(\sigma_l)$, the strange $(\sigma_s)$, and the charm $(\sigma_c)$ quark flavor basis 
\bea
\sigma_l = \frac{1}{\sqrt{2}} \sigma_0 + \frac{1}{\sqrt{3}} \sigma_8 + \frac{1}{\sqrt{6}} \sigma_{15}, \quad\;\;
\sigma_s = \frac{1}{2} \sigma_0 - \sqrt{\frac{2}{3}} \sigma_8 + \frac{1}{2\sqrt{3}} \sigma_{15}, \quad\;\;
\sigma_c = \frac{1}{2} \sigma_0 - \frac{\sqrt{3}}{2} \sigma_{15}. \hspace*{5mm}
\eea
Furthermore, one can obtain that
\begin{eqnarray}
U(\sigma_l, \sigma_s, \sigma_c) &=& - h_l \sigma_l - h_s \sigma_s - h_c \sigma_c  + \frac{m^2\, (\sigma^2_l+\sigma^2_s+\sigma^2_c)}{2} - \frac{c\, \sigma^2_l \sigma_s \sigma_c}{4}  
+ \frac{\lambda_1\, \sigma^2_l \sigma^2_s}{2}  \nonumber \\
&+& \frac{\lambda_1 \sigma_l^2 \sigma_c^2}{2} + \frac{\lambda_1 \sigma_s^2 \sigma_c^2}{2} + \frac{(2 \lambda_1+\lambda_2)\sigma^4_l }{8}   +\frac{( \lambda_1 +\lambda_2)\sigma^4_s }{4}+ \frac{(\lambda_1+\lambda_2)\sigma^4_c}{4}.\hspace*{8mm} \label{Upotio}
\end{eqnarray}

It is noteworthy highlighting that SU($3$) pure mesonic potential can be obtained when $\sigma_c \rightarrow 0$ and the additional term $c\, \sigma^2_l \sigma_s /2\sqrt{2}$ is subtracted. The latter is stemming from the anomaly term (known as $c$-term) added to mesonic potential, Eq. (\ref{lmeson}). All parameters $m^2$, $h_l$, $h_s$, $\lambda_1$, $\lambda_2$, and $c$ have been determined at different sigma masses \cite{Lenaghan,Schaefer:2008hk}. The addition parameter $h_c$ is related to the pion- and D-meson masses by Ward identities; $h_c= \sqrt{2} f_D m_D^2 - f_\pi m_\pi^2/\sqrt{2}$. In order to evaluate the PLSM chiral condensates $\sigma_l$, $\sigma_s$, and  $\sigma_c$ and the deconfinement order-parameters $\phi$ and $\phi^*$, the real part of thermodynamic potential, Re($\Omega$), should be minimized at the saddle point; 
$\left.\partial \Omega/\partial {\sigma_l}= \partial \Omega/\partial {\sigma_s}=\partial \Omega/\partial {\sigma_c}=\partial \Omega/\partial {\phi}= \partial \Omega/\partial {\phi^*}\right|_{min} =0$.

\section{Results  \label{results}}

\begin{figure}[htb]
\centering{
\includegraphics[width=4.5cm,angle=-90]{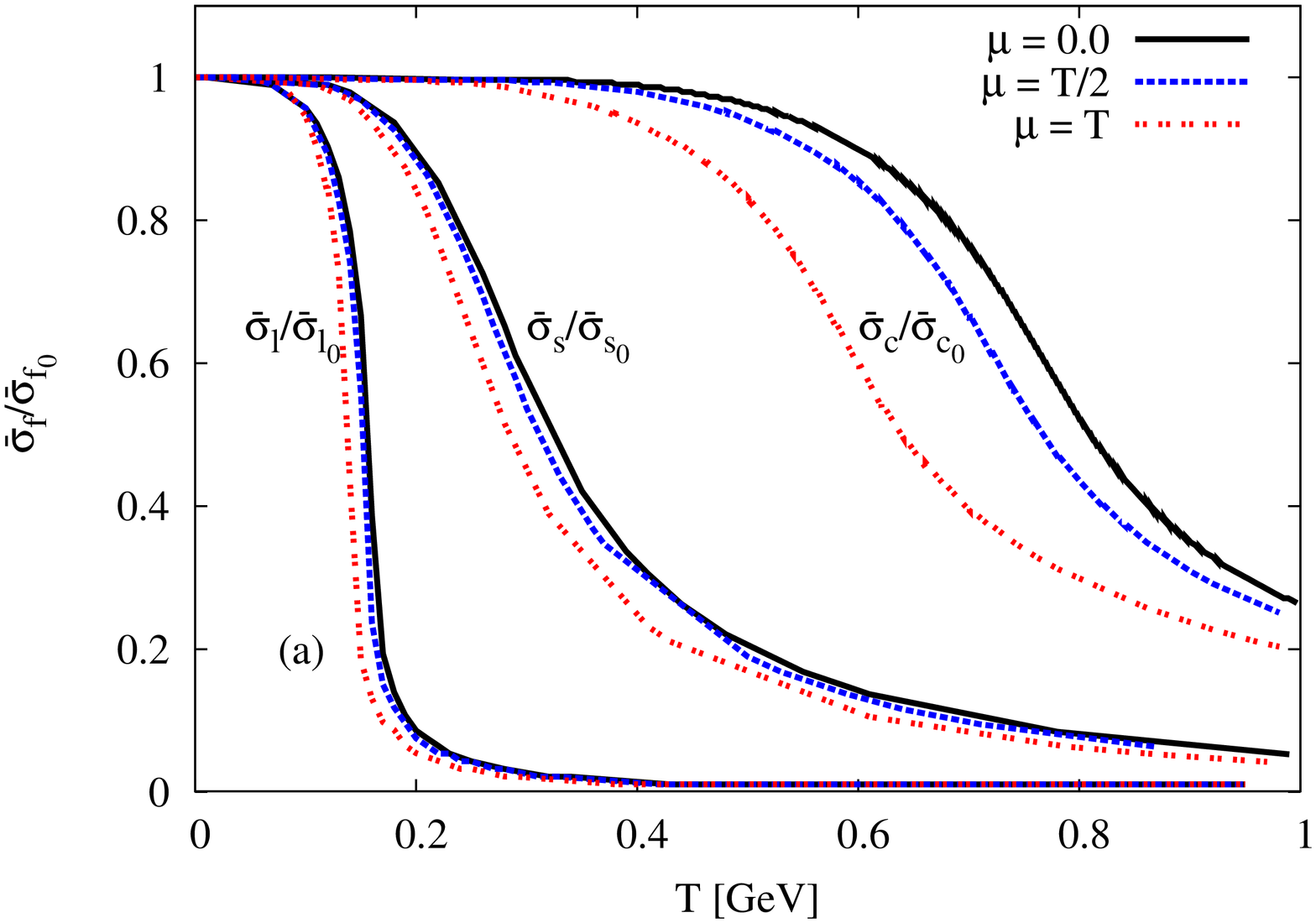}
\includegraphics[width=4.5cm,angle=-90]{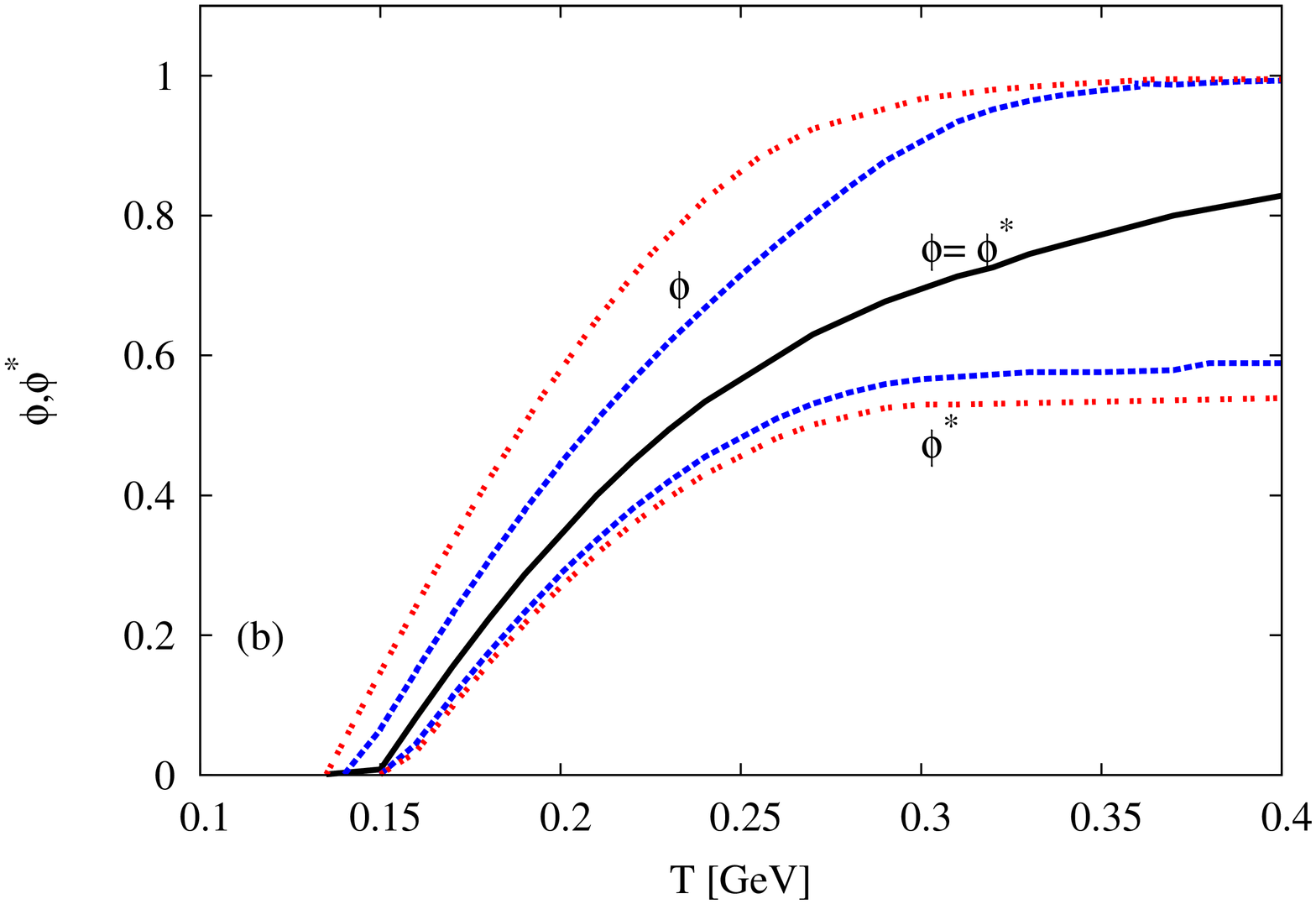}
\caption{Left-hand panel (a): normalized chiral-condensates are given as functions of temperatures at different baryon chemical potentials $\sigma_l/\sigma_{l_0}$, $\sigma_s/\sigma_{s_0}$, and $\sigma_c/\sigma_{c_0}$ (solid, dotted and double-dotted curves) for light, strange, and charm quark flavors, respectively. Right-hand panel (b):  the same as in left-hand panel but for the deconfinement order-parameters $\phi$ and $\phi^*$.
\label{cond_vis}}}
\end{figure}

The chiral condensates $\sigma_l$, $\sigma_s$, and $\sigma_c$ [left-hand panel (a)] and deconfinement order-parameters  $\phi$ and $\phi^*$ [right-hand panel (b)] have been evaluated from global minimization the real part of the thermodynamic potential at the saddle point. Left-hand panel (a) depicts the normalized chiral condensates of light, strange, and charm quarks corresponding to their vacuum values $\sigma_l=92.4$, $\sigma_s=94.5$, and $\sigma_c=295~$MeV as functions of temperatures at different baryon chemical potentials $\mu =0$, $T/2$, and $\mu=T$.  Right-hand panel (b) shows the Polyakov-loop fields  $\phi$ and $\phi^*$ estimated from the polynomial logarithmic parametrisation \cite{Sasaki:2013ssdw} at the different values of the baryon chemical potentials. It is possible to determine the chiral critical temperatures corresponding to light, strange, and charm quarks. The intersection of the deconfinement order-parameters $\phi$ and $\phi^*$ with the corresponding quark condensate at $\mu=0$ leads to $T_\chi^l = 174.6$, $T_\chi^s = 283.4$, and $T_\chi^c = 575~$MeV for light, strange, and charm quark flavors, respectively. In doing this, we recall the well-know lattice results that both chiral and deconfinement critical temperatures coincide, especially at vanishing baryon chemical potential. Alternatively, we might also apply other methods, such as, peaks of chiral condensates susceptibilities, etc. 

Also, we notice the deconfinement order-parameters $\phi$ and $\phi^*$ refer to varying critical temperatures with varying $\mu$. While the critical temperature from $\phi$ seems to decrease with increasing $\mu$, the one from $\phi^*$ increases. 

\section{Conclusion \label{Conclusion}}

In determining various PLSM parameters, the pure mesonic potential is formulated for $N_f= 4$. Accordingly, the extra degrees-of-freedom modify the thermodynamic antiquark-quark potential and the energy-momentum dispersion relations. The parameters added to the SU($4$) model, such as $\sigma_c$ and $h_c$ are found identified. We have introduced how the charm quark mass is coupled to $g$ and the charm quark condensate in vacuum. 

We present the temperature dependence of the chiral condensates $\sigma_l$, $\sigma_s$, and $\sigma_c$ and deconfinement order-parameters  $\phi$ and $\phi^*$ at varying baryon chemical potentials. We notice the critical temperatures increases when moving from light to strange to charge quark chiral condensate; $T_\chi^l = 174.6$, $T_\chi^s = 283.4$, and $T_\chi^c = 575~$MeV, respectively. This leads to draw a conclusion that the charm condensate wouldn't be affected by the well-known quark-hadron phase transition. Furthermore, we conclude that increasing $\mu$ decreases the corresponding $T_{\chi}$ and the dissociation of hadrons consisting of $c$-quarks obviously takes place at higher $T_{\chi}$. The chiral phase-structure of heavy mesons and the correlations including charm quarks in thermal and dense medium are of great importance with respect to recent experimental results.

\end{document}